\documentclass[useAMS,usenatbib,usegraphicx]{mn2e}
\pdfoutput=1

\usepackage{amsmath,amssymb,bm,upgreek,color}
\usepackage[latin1]{inputenc}
\usepackage{JournalNames}

\def\tens#1{\ensuremath{\mathsf{#1}}}

\renewcommand{\i}{\mathrm{i}}
\renewcommand{\d}{\mathrm{d}}

\newcommand{\ceff}{C_\mathrm{eff}}
\newcommand{\geff}{\gamma_\mathrm{eff}}
\newcommand{\kdv}{\bm k \cdot \delta \bm v}
\newcommand{\vA}{v_{\mathrm A}}

\newcommand{\vAphi}{v_{\mathrm A \phi}}
\newcommand{\vAz}{v_{\mathrm Az}}

\allowdisplaybreaks

\begin{document}

\title{Growth of the MRI in Accretion Discs -- the Influence of
Radiation Transport}

\author[M. Flaig et al.]{M. Flaig,$^1$
R. Kissmann,$^{1}$
W. Kley,$^1$\\
$^1$ Institut f\"ur Astronomie \& Astrophysik, Universit\"at T\"ubingen, 
Auf der Morgenstelle 10, 72076 T\"ubingen, Germany}

\date{Accepted ??. Received ??; in original form ??}

\pagerange{\pageref{firstpage}--\pageref{lastpage}} \pubyear{????}

\maketitle

\label{firstpage}

\begin{abstract}
	In this paper we investigate the influence of radiative transport on the
	growth of the magnetorotational instability (MRI) in accretion discs. The
	analysis is performed by use of analytical and numerical means. We provide
	a general dispersion relation together with the corresponding
	eigenfunctions describing the growth rates of small disturbances on a
	homogeneous background shear flow. The dispersion relation includes
	compressibility and radiative effects in the flux-limited diffusion
	approximation.  By introducing an effective speed of sound, all the effects
	of radiation transport can be subsumed into one single parameter.  It can
	be shown that the growth rates of the vertical modes -- which are the
	fastest growing ones -- are reduced by radiative transport.  For the case
	of non-vertical modes, the growth rates may instead be enhanced.  We
	quantify the effects of compressibility and radiative diffusion on the
	growth rates for the gas-pressure dominated case.  The analytical
	discussion is supplemented by numerical simulations, which are also used
	for a first investigation of the non-linear stage of the MRI in
	gas-pressure dominated accretion discs with radiation transport included.
\end{abstract}


\section{Introduction} 

Accretion discs are believed to be present in a variety of astrophysical
objects, namely active galactic nuclei, binary X-ray sources, cataclysmic
variables and protostellar systems.  The classical problem in the theory of
accretion discs is why they are accreting, or, stated in other words, what is
the mechanism which transports angular momentum outwards so that matter can
spiral inwards?  It cannot be ordinary molecular viscosity because it is
several orders of magnitude too small to account for the observed accretion
rates \citep[][]{Pringle1981ARAnA,Balbus2003ARAnA}. In order to resolve this
discrepancy, \citet{ShakuraSyunyaev1973AnA} introduced the phenomenological
$\alpha$-disc model where a large kinematic viscosity associated with turbulent
motions in the disc is thought to drive outward angular momentum transport.
What mechanism would make the discs turbulent, however, still remains a matter
of debate.  It is well known that in general shear flows are subject to
nonlinear hydrodynamical instabilities, but it seems that a compelling case for
hydrodynamical turbulence has not been made yet.  In fact, the assumption that
Keplerian discs are hydrodynamically unstable can be challenged both on
analytical and numerical grounds \citep[see,
e.g.,][]{HawleyEtAl1999ApJ,BalbusHawley2006ApJ}.  Since most accretion discs
consist of ionised gas, it is thus only logical to focus on
magnetohydrodynamical instabilities instead.  

Nowadays it is the magnetorotational instability \citep[][hereafter
BH]{BalbusHawley1991ApJ} that appears to be the most promising candidate.
The instability is local, linear, powerful (the growth rates being of the
order of the orbital period), and it exists even if the initial magnetic
field is very weak. Both three-dimensional local shearing-box
\citep[][]{HawleyEtAl1995ApJ,BrandenburgEtAl1995ApJ} and global disc
simulations \citep[][]{Armitage1998ApJ,Hawley2000ApJ,FromangNelson06}
show that the resulting turbulence indeed transports significant amounts
of angular momentum outwards, leading to values of the $\alpha$-parameter
that are in the range of $\alpha \sim 10^{-3} \cdots 10^{-2}$, depending
especially on the initial magnetic field configuration.  When compared to
observations, it seems that numerical simulations generally tend to yield
$\alpha$ values that are too small, possibly due to numerical effects
\citep[][]{KingEtAl07}.  The situation now looks rather complicated,
since it has turned out that $\alpha$ does not only depend on physical
parameters such as viscosity and resistivity, but also on details of the
numerical setup, namely the numerical dissipation and resolution \cite[]
[]{PessahEtAl07,FromangPapaloizou07}.  Therefore it is very important to
further investigate and to quantify the impact of the various physical
and numerical parameters on the MRI-induced turbulent transport.

Despite the huge amount of numerical and analytical work that has been devoted
to exploring the MRI in accretion discs, comparatively little attention has
been paid to the issue of radiation transport.  Especially when it comes to
simulations which are global in nature or at least include vertical
stratification, there are relatively few works that include radiation transport
(like \nocite{TurnerEtAl03,Turner04} Turner~et.~al.~2003, Turner~2004 for the
radiation pressure-dominated case, \nocite{KrolikEtAl07,BlaesEtAl07}
Krolik~et.~al.~2007, Blaes~et.~al.~2007 for a disc with comparable radiation
and gas pressure, and \nocite{HiroseEtAl06} Hirose~et.~al.~2006 for the
gas-pressure dominated case).  Instead usually an isothermal equation of state
is used \citep[as, for example, in the works
of][]{MillerStone00,PapaloizouNelson03,FromangNelson06}. Yet from an isothermal
model it is not possible to derive the true temperature profile, nor can one be
sure that the resulting vertical structure is correct. If, on the other hand,
one chooses an adiabatic equation of state, but does not include cooling, then
the temperature in the disc will rise quite rapidly due to turbulent heating,
which is also not very realistic. 
Only by including radiative transfer can we hope to model the energetics of
MRI-generated turbulence correctly and achieve a quasi-steady turbulent state.
In this work, we assume that the heat generated by the turbulence 
is transported by radiative diffusion.

Yet another point is, that incorporating radiative diffusion might well yield
unexpected consequences. For example it has turned out recently that the
migration of planets in a protoplanetary disc is significantly influenced
by radiative effects \citep[see, e.g.,][]{KleyCrida2008AnA}. It is not
unreasonable to expect that radiative diffusion might have other
important influences, for example when considering the growth of
planetesimals out of dust in a turbulent disc \citep[this has been
pointed out recently in][]{Brandenburg2008PS}. For these reasons,
including radiation transport should be considered imperative on the road
towards building more realistic models of accretion discs.

The plan of our paper is as follows: In Sec.~\ref{lina} we derive the
dispersion relation for axisymmetric MRI modes in a radiative, gas-pressure
dominated accretion disc in the flux-limited diffusion approximation and
calculate the corresponding eigenfunctions and MRI growth rates.  In
Sec.~\ref{num} we describe numerical simulations of single MRI modes using two
different numerical schemes and demonstrate their excellent agreement with the
analytical prediction.  We also investigate the dependence of the saturation
level of the MRI on the strength of radiative diffusion in a series of 3D local
shearing box simulations.
We conclude in Sec.~\ref{concl}, by discussing some consequences of our
analysis and providing an outlook for further research.  In the appendix we
relax the assumption that gas pressure dominates radiation pressure and provide
a general dispersion relation, including both the radiation dominated and the
gas-pressure dominated case.

\section{Linear Analysis} \label{lina}

\subsection{Dispersion relation}

We begin our analysis by deriving the MRI dispersion relation with radiation
transport included in the one-temperature flux-limited diffusion approximation
\citep[see][]{LevermorePomraning1981ApJ}.  This means we assume that gas and
radiation are at the same temperature and that the radiation energy is small
compared to the internal energy of the gas, i.e. the disc is gas-pressure
dominated.

As our starting point we take the equations of ideal magnetohydrodynamics,
namely the continuity equation
\begin{subequations}
\begin{gather}
  \label{continuity}
  \frac{\partial \rho}{\partial t} 
  +	\nabla \cdot ( \rho \bm v ) 
  = 
  0
,	
\end{gather}
the equation of motion 
\begin{equation} \label{momentum}
  \frac{\partial \bm v}{\partial t} 
  +   \bm v \cdot \nabla \bm v 
  +   \frac{1}{\rho} \, \nabla 
  \left( 
  p 
  +	\frac{B^2}{2 \mu_0} 
  \right) 
  -   \frac{\bm B \cdot \nabla \bm B}{\mu_0 \rho}
  +   \nabla \varPhi 
  = 
  0
,	
\end{equation}
and the induction equation 
\begin{gather}
  \label{induction}
  \frac{\partial \bm B}{\partial t} 
  =
  \nabla \times (\bm v \times \bm B) 
.	
\end{gather}
\end{subequations}
In Eq.~\eqref{momentum}, $\varPhi$ denotes the static gravitational potential
of the central object.  It vanishes when linearising the above equations, so we
do not need to specify it.  The other symbols have their usual meaning.

In order to perform a local stability analysis, we consider a patch of an
accretion disc that is small enough so that the background density $\rho$,
pressure $p$ and magnetic field $\bm B = B_\phi \, \hat{\bm \phi} + B_z \,
\hat{\bm z}$ can be taken constant.  We do not include a radial component in
the magnetic field; otherwise the background solution would not be time
independent because an initially radial field would be sheared into a
time-dependent azimuthal field.  For a Keplerian disc, the background flow is
locally $\bm v = -\frac 32 \varOmega r \, \hat{\bm \phi}$, with $\varOmega$ the
angular orbital frequency.  We impose plane-wave axisymmetric perturbations
$\delta \rho, \delta \bm v, \delta \bm B, \delta p$ proportional to
$\exp[\i(k_r r + k_z z ) + \sigma t]$, yielding the following linearised
equations:
\begin{subequations} \label{lin}
  \begin{align}
    \sigma \, \delta \rho 
    &=
    - 	\rho \, \i \bm k \cdot \delta \bm v
    ,	
    \label{lincon}
    \\
    \sigma \,  \delta \bm v  
    -   \frac32 \varOmega\, \delta v_r \, \hat{\bm \phi}
    &+	\frac{\i \bm k}{\rho} \cdot 
    \left( 
    \ceff^2 \, \delta \rho + \frac{\delta \bm B \cdot \bm B}{\mu_0} 
    \right)
    \notag \\
    &- 	\frac{\i k_z B_z}{\mu_0 \rho} \, \delta \bm B
    +	2 \varOmega \, \hat{\bm z} \times \delta \bm v 
    = 
    0
    ,	
    \label{linmom}
    \\
    \sigma \, \delta \bm B 
    =
    -\i \kdv \, \bm B 
    &+ \i k_z B_z
    \left(
    \delta \bm v
    -  	\frac{3 \varOmega}{2 \sigma} \, \delta v_r \, \hat{\bm \phi}
    \right)
    ;	
    \label{linind}	
  \end{align}
\end{subequations}
where we have introduced the \emph{effective sound speed} $\ceff$ via the
definition $\ceff^2 \equiv \delta p / \delta \rho$.  Note that
Eq.~\eqref{linind} ensures that the divergence-free constraint $\bm k \cdot
\delta \bm B = 0$ be satisfied.

After eliminating $\delta \rho$ and $\delta \bm B$ from the linearised momentum
equation~\eqref{linmom} by the use of~\eqref{lincon} and~\eqref{linind}, the
remaining system of three equations can be readily solved.  By defining a
wavenumber $K$ via 
\begin{equation} \label{defK}
	K \, \delta v_r \equiv \kdv
\end{equation}
(which means that $K$ basically constitutes a measure of the compressibility of
the perturbations), we can write the resulting dispersion relation in a form
that closely resembles the incompressible dispersion relation of BH: 
%
\begin{multline} \label{disp}
  \left(
  \tilde \sigma^2
  -	\frac{k_r K}{k^2} \sigma^2
  \right) \frac{k^2}{k_z^2} 
  \, 	\tilde \sigma^2
  +   \varOmega^2 \tilde \sigma^2
  -	4 \varOmega^2 k_z^2 \vAz^2
  \\
  -	2 \varOmega \sigma \vAphi \vAz k_z K
  =
  0
  ;
\end{multline}
where $k = |\bm k|$, $\bm v_\mathrm{A} = \bm B / \sqrt{\mu_0 \rho}$ denotes the
Alfv\'en speed and $\tilde \sigma^2 \equiv \sigma^2 + k_z^2 \vAz^2$.  The
form~\eqref{disp} of the dispersion relation explicitly demonstrates the
dependence on the compressibility.  The expression for $K$ turns out to be
\begin{equation}
  \label{K}
  K
  =
  \frac{
    \tilde \sigma^2 \sigma^2 k_r
    -	2 \varOmega \sigma k_z^3 \vAphi \vAz 
  }{ 
    \tilde \sigma^2
    ( \sigma^2 + k_z^2 \ceff^2 )
    +	\sigma^2 k_z^2 \vAphi^2	
  }
  ,
\end{equation}
where $\sigma$ is one of the roots of Eq.~\eqref{disp}.
In contrast to what one may expect from its definition in Eq.~\eqref{defK},
Eq.~\eqref{K} shows $K$ does not couple the modes
$\delta v_r$ and $\delta v_\phi$.  In the incompressible limit $\ceff
\rightarrow \infty$, we have $K=0$ and indeed recover the dispersion relation
of BH. 

The value of $\ceff$ is determined by the internal energy equation.  In the
one-temperature approximation suitable, e.g., for protoplanetary discs, it
reads:
\begin{equation}
  \label{energy}
  \frac{\partial e}{\partial t} +
  \nabla \cdot ( e \bm v )
  =
  -	p \, \nabla \cdot \bm v
  -	\nabla \cdot \bm F
  ,
\end{equation}
with the radiative flux vector $\bm F$ in the flux-limited diffusion
approximation
\begin{equation} \label{radflux}
  \bm F = 
  -	e D \frac{\nabla T}{T}	
  ,
  \quad \mathrm{with} \quad
  D
  \equiv	
  4 \frac{a T^4}e
  \frac{\lambda c}{\kappa \rho}
  .
\end{equation}
Here, $\kappa$ is the opacity, $c$ the speed of light and $\lambda$ the flux
limiter, which depends on the flux-limited diffusion model one decides to use.
In the linear growth phase, $\lambda$ is a constant, which is due to the fact
that we consider a uniform background.  This and the uniformity of the other
background quantities allows using $D$ as a parameter for our study, which will
be designated as the ``coefficient of radiative diffusion''.
Linearising~\eqref{energy} and using $p = (\gamma-1) \, e$, as well as the
relation $\delta T/T = \delta p/p - \delta \rho/\rho$ we obtain
\begin{equation}
  \label{linerg}
  \sigma \, \delta p =
  -\rho \ceff^2 \, \i \bm k \cdot \delta \bm v
  ;
\end{equation}
where $\ceff$ is given by the following expression:
\begin{equation} \label{soundspeed}
  \ceff^2
  =
  \geff \frac p\rho
  ,
  \quad \mathrm{with} \quad
  \geff
  \equiv
  \frac{\gamma +	k^2 D / \sigma }
       {1 + k^2 D / \sigma}
	   .
\end{equation}
We see that in the linear regime, the influence of radiative diffusion can be
described by introducing an effective adiabatic index $\geff$.  Without
radiative diffusion ($D = 0$), we have $\geff = \gamma$, and $\ceff$ is just the
adiabatic gas sound speed.  With $D$ increasing, the effective adiabatic index
$\geff$ is reduced towards the isothermal value $\geff = 1$ at $D=\infty$.
\begin{figure}
  \begin{center}%
    \includegraphics[width=0.49\textwidth]{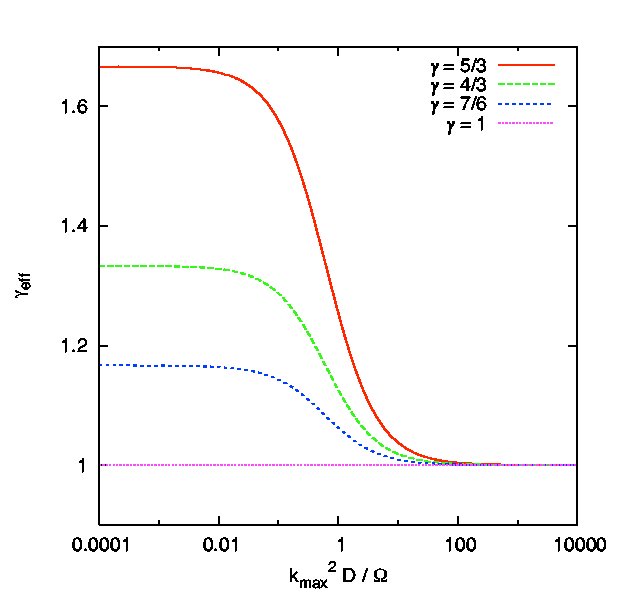}%
  \end{center}%
  \caption{\label{reduce}
  Effective adiabatic index $\gamma_\mathrm{eff}$ as a function of
  $k_\mathrm{max}^2 D / \varOmega$, where $k_\mathrm{max}$ is the wavenumber of
  the fastest growing mode.  In this plot, the azimuthal Alfv\'en speed
  is taken equal to the isothermal sound speed, $\vAphi^2 = p/\rho$ (the global features of
  this plot are not sensitive to this particular choice). 
}%
\end{figure}%
This reduction happens mostly in the regime where the nondimensional diffusion coefficient
$k^2 D / \sigma \sim 1$, as can be seen from Fig.~\ref{reduce}, where the
effective adiabatic index is plotted as a function of $D$.  Thus, in the
linear regime, the only effect of radiative diffusion is to make the physical
situation more isothermal.

In order to aid comparison 
with the dispersion relations derived by other authors, 
we define
\citep[following][]{BlaesBalbus1994ApJ,BlaesSocrates2001ApJ}
\begin{subequations}
  \begin{gather}
    \omega 
    \equiv
    \i \sigma,
    \\
    \tilde \omega^2
    =
    -	\tilde \sigma^2,
    \\
    \label{crit}
    D_\mathrm{ms}
    \equiv
    \omega^2
    - 	k_z^2 ( \ceff^2 + \vAphi^2 ),
    \\
    D_\mathrm{BH}
    \equiv
    \frac{k^2}{k_z^2} \tilde \omega^4
    -	\varOmega^2 \tilde \omega^2
    -	4 \varOmega^2 k_z^2 \vAz^2;
  \end{gather}
\end{subequations}
and write Eq.~\eqref{disp} in an alternative way as
\begin{equation} \label{dispBS}
  D_\mathrm{ms} D_\mathrm{BH}
  -	k_z^2 \vAphi^2 \vAz^2 
    ( k^2 \tilde \omega^2 + 3 k_z^2 \varOmega^2 )
    -	\frac{k_r^2}{k_z^2} \tilde \omega^2 \omega^4
  =
  0.
\end{equation}              
In the appropriate limits our dispersion relation is consistent with
that of other authors, like for example the two fluid dispersion relation of
\citet{BlaesBalbus1994ApJ} in the limit of $D \rightarrow 0$ and zero
coupling, or the resistive dispersion relation of 
\citet{SanoMiyma99} in the limit $D \rightarrow 0$
and zero resistvity.
     
\citet{BlaesSocrates2001ApJ} considered the case of a radiaton-dominated disc
where matter and radiation interact only by
momentum exchange and are, therefore, at different temperatures, while we
consider the opposite limit where matter and radiation are closely coupled via
emission and absorption processes, and are therefore at the same temperature.
In 
the appendix a more general dispersion relation is provided,
from which both the \citeauthor{BlaesSocrates2001ApJ} dispersion relation and
the one given in Eq. (\ref{disp}) can be derived.
\subsection{Eigenfunctions}
The corresponding eigenfunctions in terms of the radial velocity
perturbation $\delta v_r$ are given by the following formulae:
\begin{subequations} \label{eigen}
  \begin{gather}
    \delta \rho 
    =
    -\i \rho \frac{K}{\sigma}
    \cdot \delta v_r
    ,
    \\
    \delta v_\phi
    =	
    \frac{\sigma}{2 \varOmega k_z^2}
    \left(
    k^2 \frac{\tilde \sigma^2}{\sigma^2}
    -	k_r K
    \right) \cdot \delta v_r
    ,
    \\
    \delta v_z
    =
    \frac{K-k_r}{k_z} \cdot \delta v_r
	\label{eigen-dvz}
    ,
    \\
    \delta p
    =
    -\i \rho \ceff^2 \frac{K}{\sigma}
    \cdot \delta v_r
    ,
    \\
    \delta B_r
    =
    \i \frac{k_z B_z}{\sigma} \cdot \delta v_r
    ,
    \\
    \delta B_\phi
    =
    \i \left\{
    \frac{k_z B_z}{2 \varOmega}
    \left(
    \frac{k^2}{k_z^2} \frac{\tilde \sigma^2}{\sigma^2}
    - 	\frac{k_r K}{k_z^2}
    -	3 \frac{\varOmega^2}{\sigma^2}
    \right)
    -	\frac{K B_\phi}{\sigma}
    \right\}
    \cdot \delta v_r
    ,
    \\
    \delta B_z
    =
    -\i \frac{k_r B_z}{\sigma} \cdot \delta v_r
    .
  \end{gather}
\end{subequations}
Using once more the relation $\delta T/T = \delta p/p - \delta \rho/\rho$, we
can derive the following expression for the temperature perturbation:
\begin{equation} \label{temp}
	\frac{\delta T}{T}
  =
  \frac{\gamma - 1}{1 + k^2 D/\sigma}
  \frac{\delta \rho}{\rho}
  .	
\end{equation}
As said, radiative diffusion tends to make the situation more isothermal, and
thus for $D \rightarrow \infty$ we have $\delta T \rightarrow 0$, as is to
be expected.

\subsection{Change of growth rates due to compressibility and radiative
diffusion} \label{change}

In order to quantify the effects of a finite compressibility and radiative
diffusion on the growth rates, we first look at the vertical modes ($k_r = 0$)
which have the largest growth rates.  By differentiating the dispersion
relation~\eqref{disp} with respect to $\ceff^{-2}$, we find for the change of
the growth rate $\sigma$ with respect to a change in the effective sound speed:
\begin{equation} \label{depCeff-vert}
  \frac{1}{\sigma}
  \left( \frac{\d \sigma}{\d \ceff^{-2}} \right)_{\ceff=\infty}
  =
  -   \frac{2 k^2 \vAz^2 / \tilde \sigma^2}
  {1 + 2 \tilde \sigma^2 / \varOmega^2}
  \,	\vAphi^2
  .
\end{equation}
From this result we conclude that for vertical modes the effect of a nonzero
compressibility leads to a decrease of the growth rates in the presence of a
nonzero azimuthal field. For the fastest growing mode, which, for $\ceff =
\infty$, has a growth rate $\sigma_\mathrm{max} = 3/4 \, \varOmega$ with
wavenumber $k_\mathrm{max} = (\sqrt{15}/4) \, \varOmega / \vAz$, this means
that the change $\Delta \sigma_\mathrm{max}$ in the growth rate as compared to
the incompressible case $\ceff = \infty$ is approximately:
\begin{equation} \label{Dsmax-vert}
  \frac{\Delta \sigma_\mathrm{max}}{\sigma_\mathrm{max}}
  \approx
  -	\frac15
  \frac{\vAphi^2}{\ceff^2}
  .
\end{equation}
The relative dampening of the growth rates is, thus, in this case of
the order of $O(\vAphi^2 / \ceff^2)$.  

Curiously, if we consider non-vertical modes ($k_r \neq 0$) and a
vanishing azimuthal field, we discover the opposite. In this case we
find by an analogous analysis:
\begin{equation} \label{depCeff-nvert}
  \frac{1}{\sigma}
  \left( \frac{\d \sigma}{\d \ceff^{-2}} \right)_{\ceff=\infty}
  =
  \frac{k_r^2}{2 k_z^2}
  \frac{\tilde \sigma^2 \sigma^2 / k_z^2 \varOmega^2}
       {1 + 2 k^2 \tilde \sigma^2 / k_z^2 \varOmega^2}
       .
\end{equation}
The corresponding shift of the maximum growth rate becomes:
\begin{equation}
  \frac{\Delta \sigma_\mathrm{max}}{\sigma_\mathrm{max}}
  \approx
  \frac{27 k_r^2}{240 k^2}
  \frac{\vA^2}{\ceff^2}
  .
  \label{Dsmax-nvert}
\end{equation}
This effect
should not be considered too important, 
because the growth of the MRI
will be dominated by the fastest growing modes, which are the vertical
ones.  In both cases considered, radiative diffusion contributes a
fraction of 
$(\gamma p / \rho) / C_\mathrm{eff}^2 - 1 = \gamma/\geff - 1$ 
to the total shift in the growth rates. 

The critical wavelength, which is obtained by setting $\sigma = 0$ in
the dispersion relation, is the same as in the non-radiative case.
This means that radiative diffusion will generally not affect the
regime in which the MRI may possibly operate.  One should bear in mind
however, that not the linear analysis, but only numerical simulations,
can provide us with the answers if, in a given situation, the turbulence
that is initiated by the MRI will be sustained for a long
period of time and how effectively it transports angular momentum.
For example, recent results show that in zero-net flux shearing box
calculations this depends critically on the value of the magnetic
Prandtl number \citep[][]{FromangEtAl2007}.

To summarise, radiative diffusion changes nothing fundamental
regarding the growth of the MRI. This is because the radiative effects
enter only via the effective adiabatic index $\geff$ and, thus, change
the compressibility, but nothing else. This statement, that the
growth of the MRI under the influence of radiative diffusion can be
characterised by an effective sound speed, remains true also if we
drop the one-temperature approximation (cf. the appendix).

In gas-pressure dominated discs, the growth rates will never be dramatically
changed by radiative diffusion, since the variation in $\ceff^2$ covers
only a factor of $\gamma$.  This is different from the situation encountered in
a radiation dominated disc, where the analog of $\ceff^2$ may vary greatly.  In
the presence of a strong enough azimuthal field, the growth rates may thus be
severely reduced \citep[cf. the appendix and the paper
of][]{TurnerEtAl2002ApJ}.

\begin{figure}%
\includegraphics[width=0.55\textwidth]{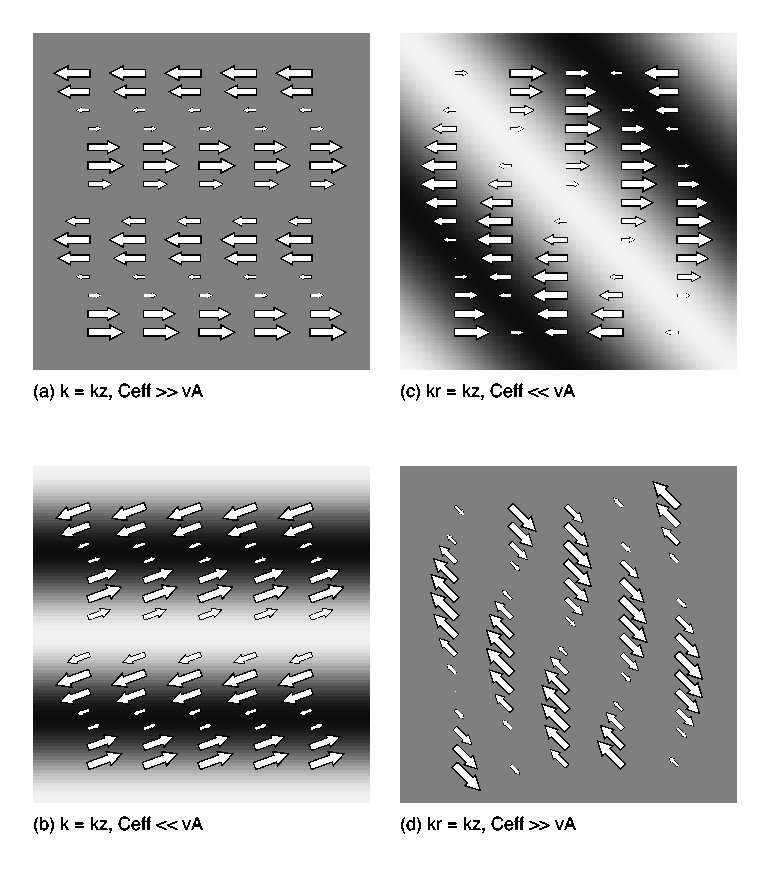}%
\caption{\label{EModes} 
	Sketch of the MRI eigenmodes.  Plotted is the density perturbation $\delta
	\rho$ (grayscale) and the velocity perturbation $\delta \bm v$ (arrows) in
	the $r$-$z$ plane.  The plots on the left show the case of a vertical mode
	in the presence of a background azimuthal magnetic field; where (a)
	corresponds to the incompressible case, while (b) is for finite
	compressibility.  The plots on the right show the case of a nonvertical
	mode with $k_r=k_z$ where no azimuthal magnetic field is present.  Here, (c)
	corresponds to the case of maximum compressibility and (d) to the
	incompressible case.
}
\end{figure}%

In order to understand the change of the growth rates due to radiative
diffusion in a qualitative manner, let us consider two cases: First, the case
of a vertical mode ($\bm k = k_z \bm e_z$) in the presence of an azimuthal
field.  In the incompressible limit ($C_\mathrm{eff} \rightarrow \infty$), the
motion of the fluid is confined to the plane perpendicular to the perturbation
wavevector, so that $\delta v_z = 0$ (Fig.~\ref{EModes} (a)).  When the
compressibility is nonzero, the Lorentz force due to the azimuthal magnetic
field causes a fluid flow in the vertical direction (Fig.~\ref{EModes} (b)).
The higher the compressibility (and the smaller therefore the gas pressure),
the more vertical will the velocity vector become.  This in turn makes the
buildup of the magnetic field less effective, resulting in a smaller growth
rate.  Radiative diffusion increases the compressibility, and therefore
decreases the growth rate.  

Next consider the case of a nonvertical mode ($k_r \neq 0$), with no background
azimuthal field.  If we now start with the limit of maximum compressibility
($C_\mathrm{eff} = 0$), we have $\delta v_z = 0$ (Fig.~\ref{EModes} (c)), since
the $z$-component of the Lorentz force vanishes.  When now increasing
$C_\mathrm{eff}$, the gas pressure will act to push the velocity vector into
the plane perpendicular to the perturbation wavevector, this effect becoming
stronger and stronger as the compressibility is further decreased
(Fig.~\ref{EModes} (d)).   Therefore, this time \textit{increasing} the
compressibility makes the buildup of the magnetic field more effective, which
means that the growth rate increases due to radiative diffusion.  

When considering the general case of a nonvertical mode in the presence of a
nonzero azimuthal field, both effects are present and the result will be either
an increase or a decrease of the growth rate, depending on the strength of the
azimuthal field and the direction of the perturbation wavevector.

\section{Numerical simulations} \label{num} 

In this section we investigate the growth of the MRI using the appropriate
numerical tools. Here we especially aim at a verification of these numerical
tools by a comparison to the analytical results derived above. After discussing
the numerical scheme we will present the results of the numerical simulations
of the MRI.  

\subsection{Numerical approach} 

The investigation of the analytical results using a numerical method is clearly
not very demanding in this case owing to the fact that the analytical model is
restricted to the linear growth phase. Here we decided, however, for a scheme,
which is well adapted to high Mach-number turbulence simulations, because we
are also interested in more general accretion disc simulations including
radiation transport. One of the most important constraints for a numerical
scheme used for compressible turbulence simulations is the correct handling of
sharp discontinuities. Here we use a second order finite volume scheme based on
the work by \citet{KurganovNoellePetrova2001} for the hyperbolic part of the
system of equations. This is a central conservative scheme for the solution of
equations of the form:
\begin{equation} \frac{\partial \bm{u}}{\partial t} + \nabla \cdot
	\tens{F}(\bm{u}) = 0.  \end{equation}
It does not require a Riemann solver and a characteristic decomposition. This
scheme was extended by a constrained transport description for the magnetic
field \citep[see, e.g.,][]{BalsaraSpicer1999JcP, LondrilloDelZanna2000} by
which we assure the solenoidality of the magnetic field. This method uses the
hyperbolic fluxes to compute electric fields on a staggered grid. These are
then used to evolve the magnetic induction, the components of which are also
given on a (different) staggered grid. The stability of the code and its
capability to resolve steep gradients without introducing artificial
oscillations have been proven, e.g., in \citet{Kissmann2006}.

Being interested in the evolution of the gas under the influence of radiation
transport we also have to deal with the corresponding source terms to the
energy equation. These were implemented using two alternative approaches. In
the first approach we included the diffusive radiation transport explicitly in
the scheme, whereas we used an implicit solver for the second method. This dual
approach is motivated by the fact that we are interested in being able to
cross-check the two different methods.

For the implicit implementation of the source terms we split the
evolution of the radiation from the evolution of the hyperbolic part
of the system of equations:
\begin{align}
  e^{(*)} 
  &=
  e^{(n)} + \left(\frac{\partial e}{\partial t}\right)_\mathrm{hyp}
  \\
  e^{(n+1)} 
  &= 
  e^{(*)} + \left(\frac{\partial e}{\partial t}\right)_\mathrm{rad}
\end{align}
where we evolve the radiation transport step using a parallel $\omega$-Jacobi
matrix solver. The $\omega$-Jacobi method is very similar to the method of
successive overrelaxation (SOR). It has the advantage that it is simple to
implement and very easy to parallelise. We have also a multigrid method
available, in case that the $\omega$-Jacobi solver should prove insufficient
for high-resolution simulations.

Having an implicit solver for the radiation transport is of special importance
with regard to future applications.  That is for high resolution simulations
with strong radiation transport the time-step of an explicit scheme would
become prohibitively small, whereas an implicit solver would only be limited by
the Courant condition \citep[see][]{CourantFriedrichLewy1928} for the
hyperbolic part of the system.

Our second implementation is nonetheless of explicit form, because we wanted to
have a completely different method at hand to be able to compare it to the
implicit scheme for more general simulations for which no analytical solution
is available. For this purpose we extended the flux-function for the energy
equation by:
\begin{equation} \label{implicit}
  \bm{F}^\mathrm{Rad} 
  =
  -\frac{e}{T} D \left(\frac{\partial T}{\partial x} \bm{\mathrm e}_x +
  \frac{\partial T}{\partial y} \bm{\mathrm e}_y +
  \frac{\partial T}{\partial z} \bm{\mathrm e}_z\right)
\end{equation}
Since the flux functions are only needed at the cell faces the occurring
gradients are easily computed as can be seen from the example of the
$x$-component:
\begin{equation}
  F^\mathrm{Rad}_{x,i+1/2,j,k} 
  =
  -\frac{e}{T} D 
  \left(
  \frac{T_{i+1,j,k} - T_{i,j,k}}{\Delta x}
  \right)
\end{equation}
This scheme proved to be stable for:
\begin{equation}
  \Delta t < \frac{\Delta x^2}{8 D} \frac{\rho T/e}{\gamma - 1}
\end{equation}
In the following section we will show that both implementations for
the radiation transport are well-behaved, yielding correct result when
applied to the analytical problem at hand. Thus, they can be used for
future simulations of more complex radiation transport phenomena.
\subsection{Growth rates}
\begin{figure}%
  \begin{center}%
    \includegraphics[width=0.49\textwidth]{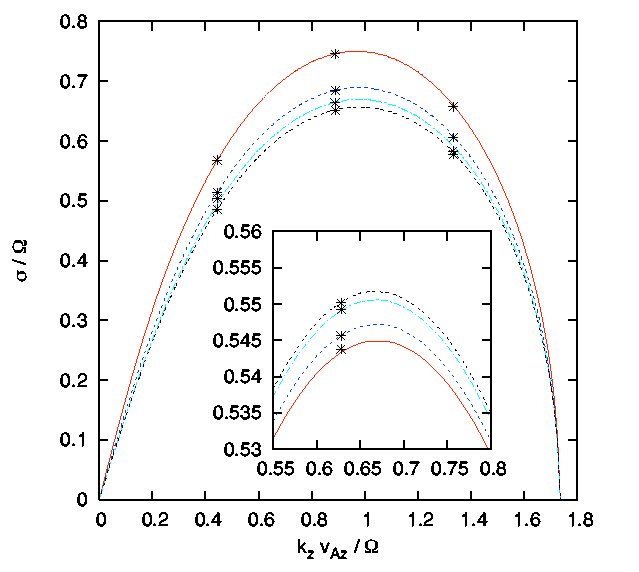}%
  \end{center}%
  \caption{\label{growkz} 
	MRI growth rates obtained with the implicit solver.  Plotted are the
	analytical solutions for various cases, together with corresponding data
	points obtained by numerical simulations.   The curves in the main part of
	the graphic are for vertical modes ($k_r=0$), where the growth rates are
	reduced by radiative diffusion.  From top to bottom, they represent the
	following cases, $B_\phi = D = 0$; $B_\phi = 20 B_z, D = 10^{-5}$; $B_\phi
	= 20 B_z, D = 10^{-4}$; $B_\phi = 20 B_z, D = 1$.  The vertical magnetic
	field in terms of the plasma beta, $\beta = 2\mu_0 p/B^2$, is chosen such
	that $\beta = 400$, which means that $v_{Az}/\sqrt{p/\rho} \simeq 0.1$.
	The inset shows simulation results for nonvertical modes with no azimuthal
	magnetic field, where radiative diffusion increases the growth rates.  We
	chose $k_r=k_z$, $B_\phi=0$ and $B_z$  such that $v_{Az}
	= \sqrt{p/\rho}$.  From bottom to top, we have $D=0$, $D=10^{-5}$,
	$D=10^{-4}$, $D=1$.  Although the difference in the growth rates is only
	small for the latter case, it is accurately reproduced by our numerical
	code.}  

\end{figure}%
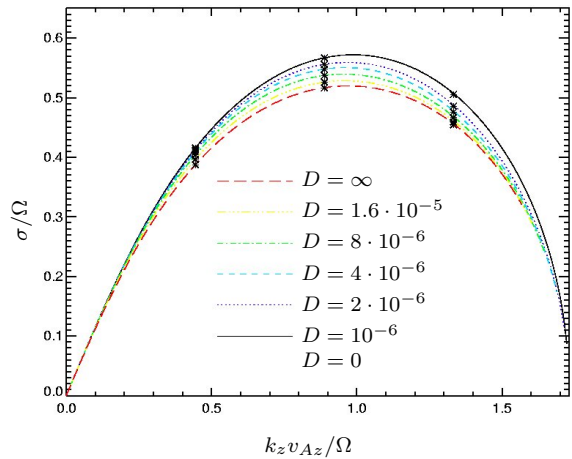
\begin{figure}
  \setlength{\unitlength}{0.0004\textwidth}
  \begin{picture}(1100,875)(-100,-100)
    \put(420,-70){$k_z v_{Az}/\Omega$}%
    \put(-60,330){\rotatebox{90}{$\sigma/\Omega$}}%
    \includegraphics[width=1000\unitlength]{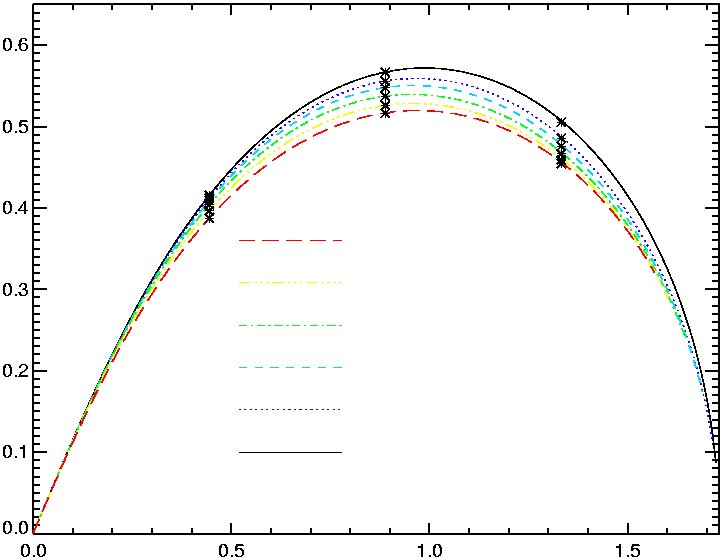}
	\put(-510,90){$D=0$}
    \put(-510,138){$D=10^{-6}$}
    \put(-510,197){$D=2\cdot 10^{-6}$}
    \put(-510,256){$D=4\cdot 10^{-6}$}
    \put(-510,315){$D=8\cdot 10^{-6}$}
    \put(-510,374){$D=1.6\cdot 10^{-5}$}
	\put(-510,433){$D=\infty$}
  \end{picture}
  \caption{\label{FigGrowthExlicit}Comparison of the numerical growth
  rates (black crosses) to the analytical model results (full
  lines). Here we show the data from the simulations with the explicit
  numerical implementation of the radiation transport terms. 
  All simulations have $B_\phi = 25 B_z$ and a vertical magnetic
  field corresponding to a plasma beta $\beta = 400$.  The curve
  for $D=\infty$ has been inserted to show the maximum possible
  change in the growth rates.
}
\end{figure}

For our numerical simulations, we use the shearing box approximation
\citep[see, e.g.,][]{HawleyEtAl1995ApJ}.  In order to compare the analytical
growth rates of the MRI to the results from our numerical code, we perform
simulations of single MRI eigenmodes.  Since only variations in the radial and
in the $z$-directions have to be taken into account, it is sufficient to use a
2D model. We set up the problem using the eigenfunctions as given
by~\eqref{eigen}.  The size of the computational domain is [-0.5,0.5] x
[-0.5,0.5] ($x$- and $z$-direction respectively) with a typical resolution of
64x256 cells. Initially we set $\varOmega = 10^{-3}$, density $\rho = 1$ and
pressure $p = 10^{-6}$.
The simulations were
performed for different wavenumbers and for various strengths of 
magnetic field and radiative diffusion coefficient
(for the explicit values see in the figures).

The results are plotted in Figs.~\ref{growkz} and
\ref{FigGrowthExlicit} for the implicit and the explicit
implementation of the source terms, respectively, together with the
corresponding analytical solution.  
In the simulations with the 
explicit scheme a constant diffusion coefficient was used
[cf. Eq.~\eqref{implicit}],
while in the simulations with the implicit scheme, the full radiation 
transport was done as prescribed by Eq.~\eqref{radflux}, with
$\lambda = 1/3$ and constant opacity $\kappa $.  
The data points nicely match the
analytical prediction, thereby confirming both the validity of the
numerical scheme and the analytical calculation. In particular it is
evident that both implementations of the radiation transport are in
good agreement to the analytical predictions. Therefore, we can use
both methods to simulate more complex situations in the future giving
us the good opportunity to have two different methods to be applied to
the same problem. Thus, it will be easier to decide if observed
effects might be due to some numerical problem.

\subsection{Saturation level}

In order to study the impact of radiation transport on the saturation level of
the MRI, we switch to a 3D representation of the shearing box with a
computational domain [-0.5,0.5] x [0,4] x [-0.5,0.5].  Concerning the values of
$\rho,p,\varOmega$ we use a similar setup as in the 2D simulations described in
the previous section.  We choose an initially vertical magnetic field
corresponding to a plasma beta of $\beta = 400$ in all simulations.  Instead of
prescribing eigenfunctions, we initialise the problem with random velocity and
pressure perturbations of order $10^{-6}$.              

The strength of the angular momentum transport is described by the 
$\alpha$-parameter \citep{ShakuraSyunyaev1973AnA}.  In our simulations, it
is measured according to the following prescription:
\begin{equation}
	\alpha = \langle T_\mathrm{R} + T_\mathrm{M}\rangle /p_0,
\end{equation}
where $p_0$ is the initial gas pressure, $T_\mathrm{R} = \rho \delta v_r \delta
v_\phi$ denotes the Reynolds stress and $T_\mathrm{M} = - \delta B_r \delta B_\phi$ is the
Maxwell stress.  The angular brackets $\langle \cdots \rangle$ indicate a
spatial average.  We also use the dimensionless stresses 
$\alpha_\mathrm{R}$ and $\alpha_\mathrm{M}$ defined as
\begin{equation}
	\alpha_\mathrm{R} = \langle T_\mathrm{R} \rangle / p_0,
	\quad
	\alpha_\mathrm{M} = \langle T_\mathrm{M} \rangle / p_0.
\end{equation}
Note that with this definitions in general we will have $\alpha \neq
\alpha_\mathrm{R} + \alpha_\mathrm{M}$.  Due to the fact that in our 
simulations the magnetic field posesses a net flux, the resulting $\alpha$
is quite large, $\alpha \gtrsim 0.1$. 

\begin{figure*}
  \setlength{\unitlength}{0.00044\textwidth}
  \begin{picture}(1100,1100)(-100,-100)
    \put(-60,400){\rotatebox{90}{$E_\mathrm{mag}/p_0$}}%
    \put(-100,-100){\includegraphics[width=1200\unitlength]{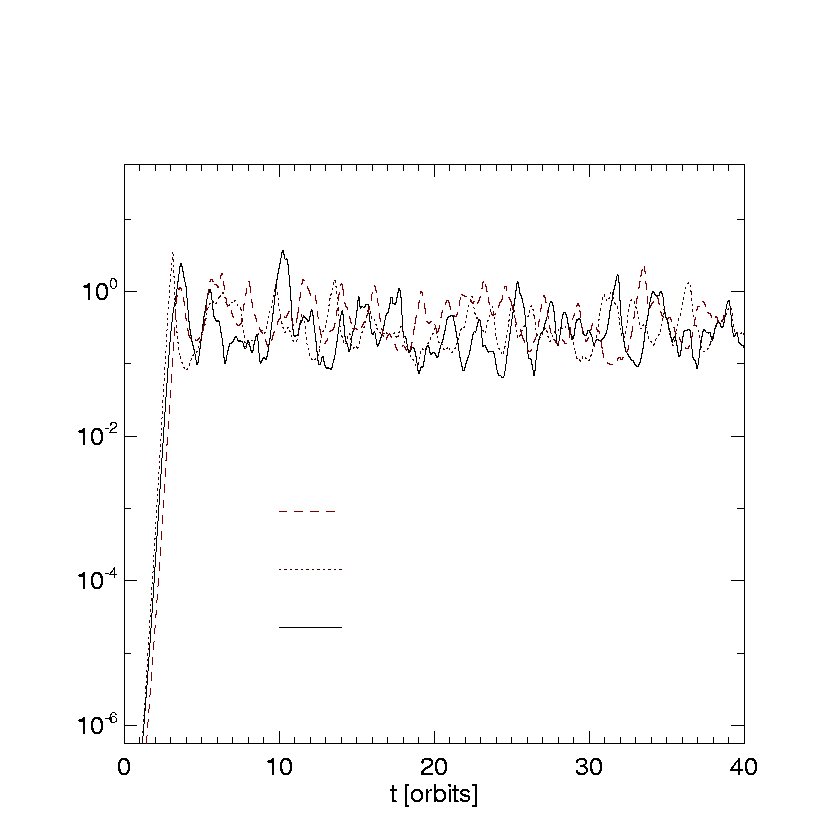}}%
    \put(420,350){32x64x32}%
    \put(420,270){64x128x64}%
    \put(420,180){128x256x128}%
  \end{picture}
  \hfill
  \begin{picture}(1100,787)(-100,-100)
    \put(-40,450){\rotatebox{90}{$\alpha$}}%
    \put(-100,-100){\includegraphics[width=1200\unitlength]{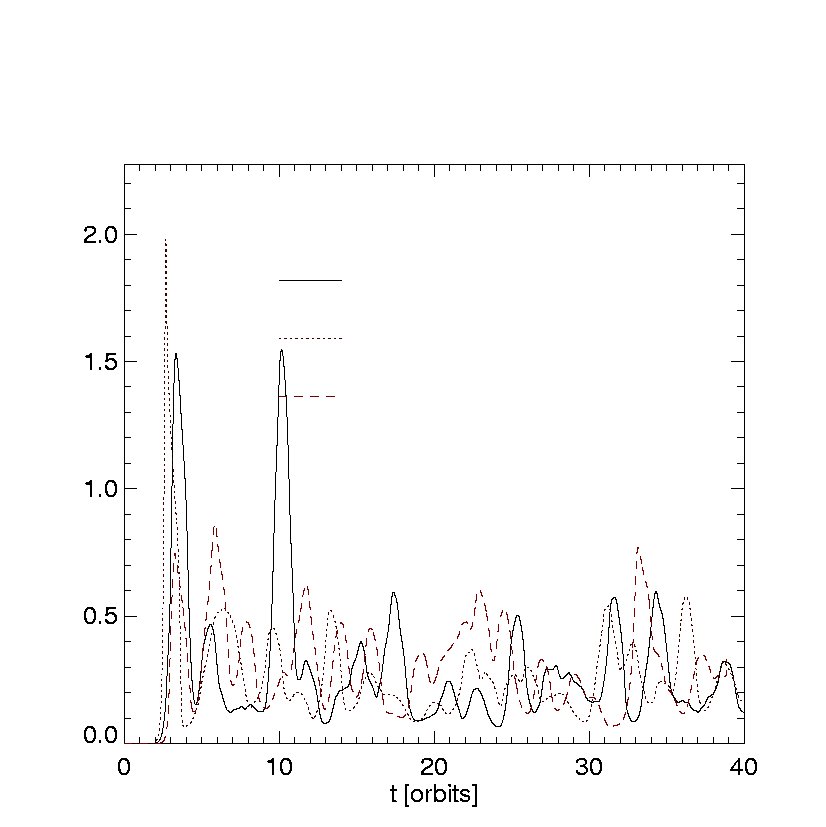}}%
    \put(420,690){32x64x32}%
    \put(420,600){64x128x64}%
    \put(420,520){128x256x128}%
  \end{picture}
  \caption{\label{Stress}
	Resolution dependence of the turbulent activity on the numerical
	resolution.  The left and right panels show the time evolution of the
	perturbed magnetic energy and the alpha parameter, respectively.  In case
	of the alpha parameter the curves have been smoothed over 1 orbit.
	When considering long-term averages of these quantities, there
	appears to be no trend with varying resolution.}
\end{figure*}
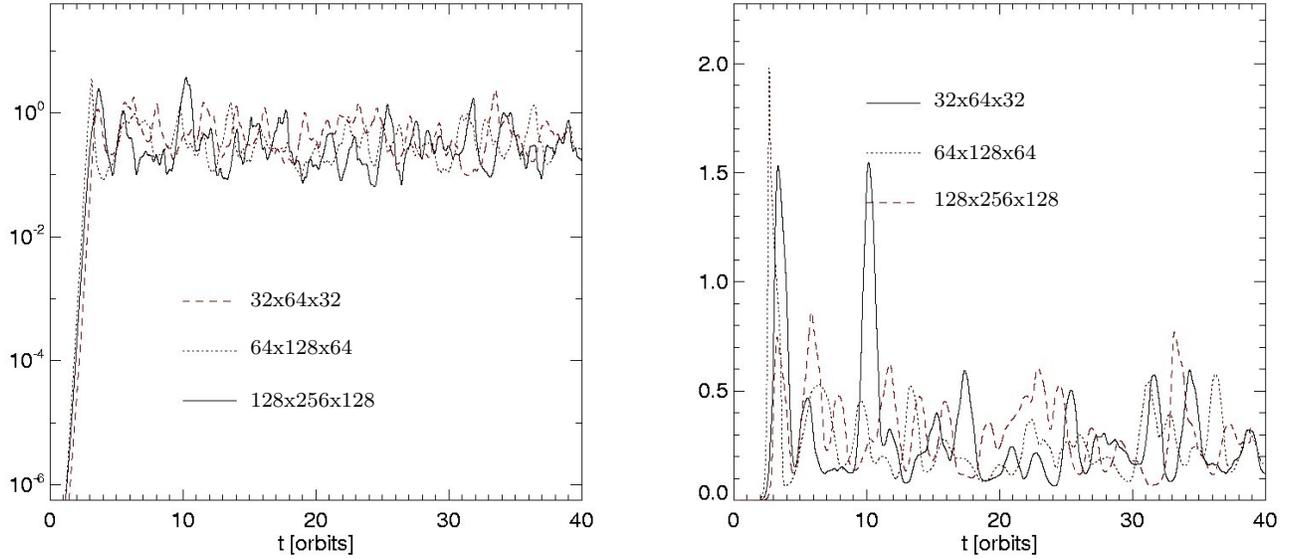

In order to check if there exists a trend for the turbulent activity with
different resolution, we performed three runs where the resolution was
successively doubled.  Fig.~\ref{Stress} shows the time evolution of the
magnetic energy and the $\alpha$-parameter for this runs.  In
Table~\ref{cronos} long-term averages of this quantities as well as of the
Maxwell and Reynold stresses can be found.  The
results displayed in Fig.~\ref{Stress} and Tab.~\ref{cronos} indicate no net trend
for the change of the saturation level with
resolution (the first three lines in the table).

As we have seen, the analytical calculation predicts that radiative diffusion
modifies the MRI growth rates.  The question if the saturation level of 
the MRI-induced turbulence is also changed by radiative effects can only 
be answered by numerical simulations.  
We have performed simulations with varying radiative diffusion coefficient,
where we picked a resolution of 64x128x64.  The results show a 
moderate decrease of the turbulent activity with inreasing radiative diffusion coefficient $D$,
see last 3 entries in Tab.~\ref{cronos}.  
A qualitative explanation for this is given below.


\begin{table} 
\caption{Time- and space-averaged quantities from 3D shearing box simulations
performed with our code.  In all of the simulations $B_\phi = 0$.  In the first
three lines, only the resolution is varied, while the last three lines show the
dependence on the strength of radiative diffusion.  Averages were taken over 10
to 40 orbits.
\label{cronos}
}
\begin{tabular}{cccccc}
\hline
Resolution	& $D$  & $E_\mathrm{mag} / p_0$ & $\alpha$ & $\alpha_\mathrm{M}$ & 
$\alpha_\mathrm{R}$ \\ \hline
32x64x32    & 1	& 0.400	& 0.278	& 0.248 & 0.044  \\
128x256x128 & 1	& 0.505	& 0.299	& 0.279 & 0.036  \\ 
64x128x64	& 1		& 0.337	& 0.226	& 0.195	& 0.031	\\
\hline
64x128x64	& 0 & 1.007	& 0.719	& 0.599	& 0.119	\\
64x128x64	& $10^{-5}$	& 0.358	& 0.244	& 0.210	& 0.034	\\
64x128x64	& 1		& 0.337	& 0.226	& 0.195	& 0.031	\\
\hline
\end{tabular}
\end{table}

\begin{figure}%
  \begin{center}%
    \includegraphics[width=0.49\textwidth]{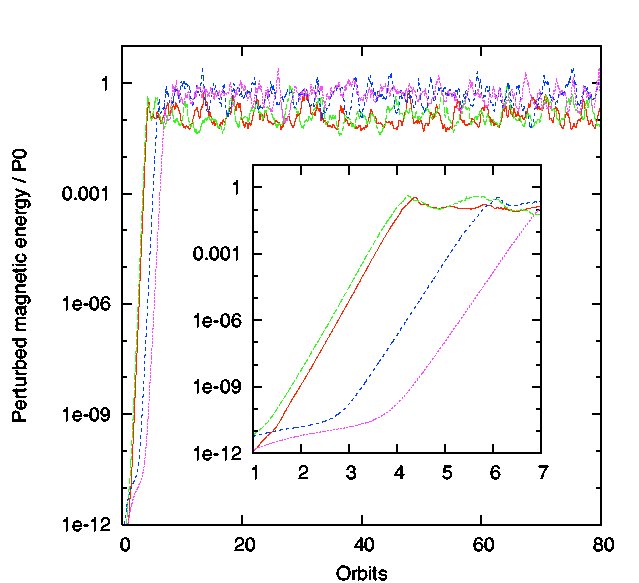}%
  \end{center}%
  \caption{\label{sl64sat} 3D shearing box simulations performed with ZEUS at
  resolution 64x128x64.  Shown is
  the perturbed magnetic energy normalised to the initial gas pressure. 
  The solid, dashed, dot-dashed and dotted curves correspond to the combinations
    $B_\phi = 0$/$D = 0$, $B_\phi = 0$/$D=1$, 
    $B_\phi = 20 B_z$/$D=0$ and $B_\phi=20 B_z$/$D = 1$, respectively.
In the linear phase, the obtained growth rates are only a few
percent smaller than the growth rate of the fastest growing mode
that fits in the box, which means that the initial growth is
dominated by the fastest growing mode, as is to be expected.}
\end{figure}

We have also performed additional simulations with the ZEUS code in order to
check this result using a different numerical scheme.  We used the same setup
concerning $\rho$,$p$,$\varOmega$ as in the simulations done with our code.
The strength of the constant vertical magnetic field was taken as corresponding
to a plasma beta of $\beta = 800$.  We performed simulations with and without
azimuthal magnetic field $B_\phi$, using different resolutions (see
Table~\ref{zeus}).  In Fig.~\ref{sl64sat}, the magnetic energy is plotted for
the high resolution simulations.  As has already been stated in the
introduction, it is known that the outcome of an MRI simulation does depend on
numerical issues such as the resolution or the numerical magnetic Prandtl
number of the code~\citep{FromangPapaloizou07}.  Indeed, if we compare for
example the simulations in lines 3 and 4 of Tab.~\ref{cronos} with the
corresponding simulations done with the ZEUS code, lines 5 and 6 of
Tab.~\ref{zeus}, the ratio of the saturation levels in the case of the ZEUS
code turns out to be about a factor of two smaller than in the simulations
done with our code.  We have checked that this fact remains true even if we
choose the same initial plasma beta as in the runs with our code.  Despite
these issues, all runs unambigously show that the saturation level decreases
with increasing radiative diffusion.

\begin{table}
\caption{3D shearing-box simulations performed with ZEUS.
In some of the simulations, a non-zero constant $B_\phi$ was used,
leading to a higher saturation level.
Averages were taken from 10 to 40 orbits.
\label{zeus}
}
\begin{tabular}{ccccccc}
\hline
Resolution & $D$		& $\frac{B_\phi}{B_z}$ & $E_\mathrm{mag}/p_0$ & $\alpha$ & $T_\mathrm{M}$ & $T_\mathrm{R}$\\
\hline
32x64x32	& 0		& 0		& 0.398		& 0.248	& 0.210	& 0.038	\\
32x64x32	& 1		& 0		& 0.245		& 0.146	& 0.123	& 0.023	\\
32x64x32	& 0		& 20	& 2.557		& 1.125	& 0.983	& 0.141	\\
32x64x32	& 1		& 20	& 1.542		& 0.577	& 0.498	& 0.079	\\
64x128x64	& 0		& 0		& 0.552		& 0.335	& 0.288	& 0.047	\\
64x128x64	& 1		& 0		& 0.392		& 0.218	& 0.187	& 0.031	\\
64x128x64	& 0		& 20	& 2.474		& 1.137	& 1.006	& 0.131	\\
64x128x64	& 1		& 20	& 2.315		& 1.075	& 0.938	& 0.137	\\
\hline
\end{tabular}
\end{table}

A plausible explanation for the phenomenon of the reduction
of the saturation level due to radiative diffusion can be given as
follows:
In the case of an initial magnetic field with nonzero net flux, the turbulence 
in the saturated state is still partially 
pumped by the two-channel mode (the vertical mode with the longest 
wavelength that fits into the box).  The channel mode reappears every few orbits; 
a manifestation of this are the oscillations that can be seen in the time evolution of the 
magnetic energy and the $\alpha$-parameter \citep[see, for example][]{SanoInutsaka2001}.
In Sec.~\ref{change} we have seen that radiative transport tends to decrease
the growth rates of vertical modes by effectively increasing the compressibility
of the fluid and thus making it easier for the Lorentz force
to push the velocity vectors out of the $r$-$\phi$ plane.  Therefore 
it is to be expected that radiative transport, by decreasing the growth rate
of the recurrent channel mode, acts to reduce the turbulent
activity.
%
\subsection{Temperature distribution}
\begin{figure*}
  \setlength{\unitlength}{0.00044\textwidth}
  \begin{picture}(1100,872)(-100,-100)
    \put(420,-70){$T$}%
    \put(-60,360){\rotatebox{90}{$f_T$}}%
	\includegraphics[width=1000\unitlength]{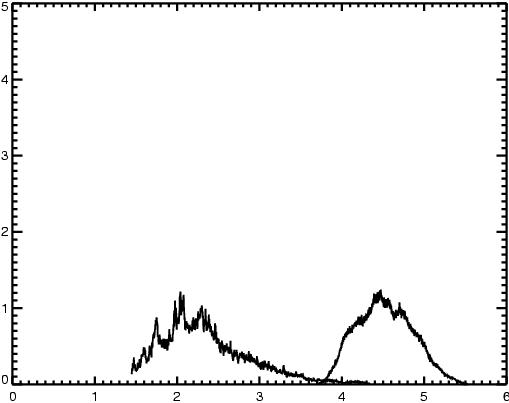}
  \end{picture}
  \hfill
  \begin{picture}(1100,872)(-100,-100)
    \put(420,-70){$T$}%
    \put(-60,360){\rotatebox{90}{$f_T$}}%
	\includegraphics[width=1000\unitlength]{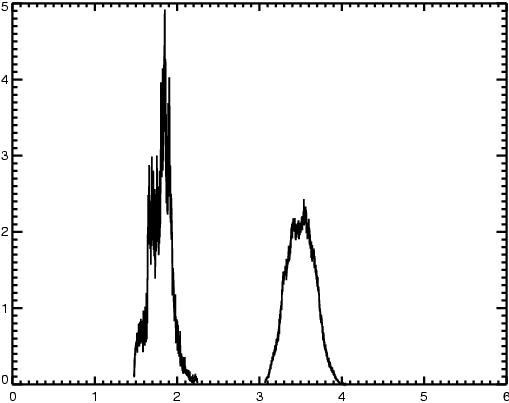}
  \end{picture}
  \caption{\label{FigTDist}Temperature distribution for two 3D
  simulations without radiation transport (left) and with radiative 
  transport, $D = 4\cdot10^{-5}$ (right).  In both plots, the distribution
  to the left corresponds to $t = 3.5$ orbits, while the right corresponds
  to $t = 4.1$ orbits.  In the case with radiative transport, the 
  width of the distribution function is smaller and the heating occurs
  less rapidly.}
\end{figure*}

\begin{figure*}
  \setlength{\unitlength}{0.00044\textwidth}
  \begin{picture}(1100,787)(-100,-100)
    \put(420,-70){$\rho$}%
    \put(-60,320){\rotatebox{90}{$e_{th}$}}%
    \includegraphics[width=1000\unitlength]{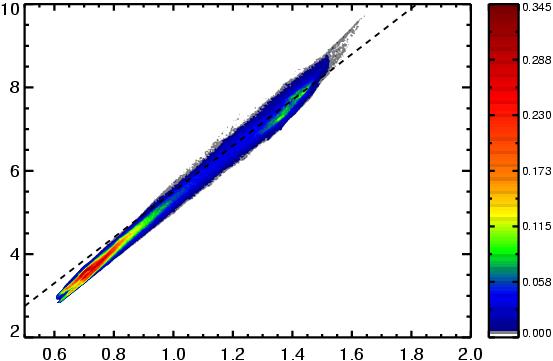}
  \end{picture}
  \hfill
  \begin{picture}(1100,787)(-100,-100)
    \put(420,-70){$\rho$}%
    \put(-60,320){\rotatebox{90}{$e_{th}$}}%
    \includegraphics[width=1000\unitlength]{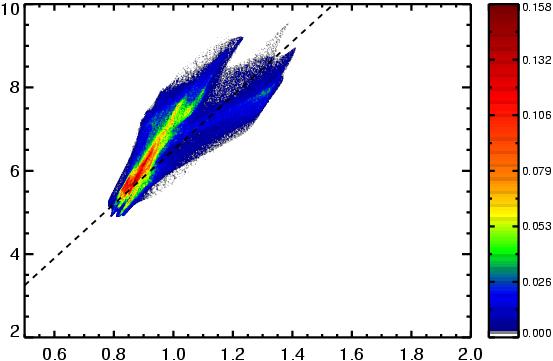}
  \end{picture}
  \caption{\label{FigRPDist}Distribution in the density- thermal energy
  plane. Values are shown in normalised units.  Here we show the
  distribution function for a simulation without radiative transport
  (left) and one where we used $D = 4\cdot10^{-5}$ (right).  Both 
  snapshots are taken at time $t = 4.1$ orbits.}
\end{figure*}

For the full 3D simulation the influence of the radiation transport
can also be visualised by the temperature distribution. This is shown
in Fig. \ref{FigTDist} for a similar setup as described in the
previous section. Here we used an extent of the computational domain
of [-0.5,0.5] x [0,6] x [-0.5,0.5] with the same spatial resolution
of 64x128x64.  The $z$-component of the magnetic induction was initialised
with $\beta = 800$, and we chose $B_\phi=20 B_z$. We followed the
evolution for several orbits.
In Fig. \ref{FigRPDist} we show the distribution in the density -
thermal energy plane for one simulation without and one with 
radiative transport.
Apparently
the plasma gets more isothermal when radiation transport is present --
for a fully isothermal simulation the thermal energy would depend
linearly on the density, thus, yielding a line in this plot. This fact
also becomes apparent, when comparing the results for the temperature
distribution function.

This is done in Fig. \ref{FigTDist} for the same simulations. Here,
the temperatures are normalised to the initial temperature (that is,
initially we had a delta-peak at $T=1$). The shift of the distribution
functions with regard to the initial temperature results from the
heating by (numerical) viscosity and resistivity.  Obviously the
distribution for the disc with radiation transport is considerably
narrower. Note, however, that also for this case the temperature
after several orbits differs markedly from the initial
temperature. This conlusion is also strengthend by the fact that both
distributions are centered at nearly the same temperature. This is due
to the fact that the gas is still heated by viscosity and
resistivity. In contrast to the case without radiation transport the
local temperature enhancements are smoothed out by the radiation
transport.
\section{Discussion} \label{concl}
We analysed the influence of radiation transport on the
magnetorotational instability. We started by investigating the linear
growth of small disturbances on an underlying constant shear flow. The
system under investigation is a magnetised shearing box, where we
explicitly retained the compressibility of the gas. Additionally we
took radiation transport via flux-limited diffusion into account. For
this system we derived a general dispersion relation, which in the
appropriate limits can be reduced to several dispersion relations
known from the literature.

Here, we placed special emphasis on the one-temperature flux-limited
diffusion approximation suited for accretion discs not dominated by
the radiation. The corresponding dispersion relation can also be
derived from the general relation given in the appendix. From this we
computed the growth rates for the linear case, which we then also
compared to the results from numerical simulations. The excellent
agreement between simulations and analytical prediction not only
confirmed the accuracy of the employed numerical schemes but also the
correctness of the analytical solution.

This analysis was then further extended by numerical simulations,
which also took the non-linear phase of the instability into
account. These simulations were done for a more general case, where
several modes of the instability where excited in a fully three
dimensional setup. There we concentrated on the resulting saturation
levels of the turbulence and on the temperature distribution, which is
also influenced by the radiation transport.

From our analysis it is apparent that the influence of the radiation
transport is a rich phenomenon. It not only weakens the
magnetorotational instability, but can also enhance it in a certain
parameter regime. This shows that radiation transport -- even in the
one-temperature flux-limited diffusion limit -- does not simply act as
a dissipative process like viscosity and resistivity. Another
difference between these processes is that the latter can completely
suppress the instability, whereas radiation transport can only weaken
it. This is due to the fact that a diffusive radiation transport only
smoothes the temperature, whereas the other dissipative processes
directly act onto either velocity or magnetic field
fluctuations. Thus, we have, for strong radiation transport,
approximately an isothermal state of the plasma. This, however, still
differs from the use of an isothermal equation of state with the
difference arising from the fact that when using an energy equation
with $\gamma$ differing from one, we still have heating due to
dissipative processes (either physical or numerical). Thus, the effect
of radiation transport is just to smooth out the local peaks of heat
appearing due to local dissipation.

This picture will probably become different, when investigating the
MRI in an open box instead of the periodic shearing box. The future
task will, thus, be the investigation of a stratified shearing box,
where we can allow for outgoing radiation in the halo of the
disc. This way even a nearly isothermal core of the disc might occur,
which might then be compared to simulations using an isothermal
equation of state. Here, however, we have restricted ourselves to
the investigation of the local shearing box, since only this can be
approached by analytical means.
\subsection*{Acknowledgements}
This research has been supported in part by the Deutsche Forschungsgemeinschaft
DFG through grant DFG Forschergruppe 759 ``The Formation of Planets: The
Critical First Growth Phase''.  Computational resources were provided 
by the High Performance Computing Cluster of the University of
T\"ubingen and in the
project hbo25 by the Forschungszentrum J\"ulich.  We thank the anonymous referee
for providing valuable suggestions that helped to improve the paper.
\appendix
\section{General dispersion relation}
\label{SecAppend}
In this appendix, we provide the general dispersion relation by
including the radiation energy equation and dropping the
one-temperature approximation.  This means that in the momentum
equation~\eqref{momentum}, we have to replace $\nabla p$ by 
$\nabla p + \lambda \nabla E$,
where $E$ denotes the radiation energy density. As in
Eq. (\ref{radflux}) $\lambda$ denotes the flux limiter, which is
constant in the linear growth phase due to the constant background. The
effect of this is that in the linearised momentum
equation~\eqref{linmom}, we should now define the effective sound
speed $\ceff$ as
\begin{equation} \label{genC}
\ceff^2 
\equiv
\frac{\delta p}{\delta \rho} 
+ 
\lambda
\frac{\delta E}{\delta \rho}
.
\end{equation}
To determine it, we need the internal energy equation and the
radiation energy equation:
\begin{subequations} \label{rad}
  \begin{gather}
    \frac{\partial e}{\partial t} +        
    \nabla \cdot ( e \bm v )                      
    =                                                               
    -	p \, \nabla \cdot \bm v                                              
    -	\kappa_\mathrm{P} \rho ( 4 \uppi B - c E )
    ,
    \\
    \label{raden}
    \frac{\partial E}{\partial t} +
    \nabla \cdot ( E \bm v )
    =
    -	\frac E3 \, \nabla \cdot \bm v
    +	\kappa_\mathrm{P} \rho ( 4 \uppi B - c E )
    -	\nabla \cdot \bm F 
    .
  \end{gather}
\end{subequations} 
Here, $B$ denotes the Planck function and $\kappa_\mathrm{P}$ the Planck
mean opacity.  The terms containing $\kappa_\mathrm{P}$ describe the
coupling of the gas to the radiation via emission and absorption.
The radiative flux vector now reads
\begin{gather}
    \bm F = 
    -	\frac{\lambda c}{\kappa_\mathrm{F} \rho}
    \nabla E
,
\end{gather}
where $\kappa_\mathrm{F}$ is the flux-mean total opacity.  We define
\begin{equation}
  \alpha_\mathrm{P}
  \equiv
  \frac{\kappa_\mathrm{P} \rho c}{\sigma}
  \quad \mbox{and} \quad
  \alpha_\mathrm{F}
  \equiv
  \frac{k^2}{\sigma} \frac{\lambda c}{\kappa_\mathrm{F} \rho},
\end{equation}
The dimensionless quantity $\alpha_\mathrm{P}$ gives the rate at which
radiation and matter are thermally coupled with respect to the growth rate of
the instability 
while $\alpha_\mathrm{F}$ constitutes a non-dimensional diffusion coefficient.
We assume that in the equilibrium the gas temperature $T_\mathrm{g}$ and
radiation temperature $T_\mathrm{r}$ defined by $E = a T_\mathrm{r}^4$ are
equal: $T_\mathrm{g} = T_\mathrm{r} \equiv T$ and $\frac{4 \uppi B}{c} = a
T^4$.                                                 Now the linearised
equations corresponding to Eqs.~\eqref{rad} become: 
\begin{gather} \label{linrad}
  \delta p
  =
  \gamma \frac p\rho \delta \rho
  -	(\gamma - 1) \alpha_\mathrm{P} E \,
  \left(
  4 \frac{\delta T_g}{T} - \frac{\delta E}{E}
  \right)
  ,
  \\
  \delta E
  =
  \frac{4 E}{3 \rho} \delta \rho
  +	\alpha_\mathrm{P} E \,
  \left(
  4 \frac{\delta T_g}{T} - \frac{\delta E}{E}
  \right)
  -	\alpha_\mathrm{F} \, \delta E
  ;
\end{gather}
From these linearised equations we find after some algebra
\begin{subequations}
	\begin{gather}
    	\frac{\delta p}{\delta \rho}
		=
		\frac{(1 + \alpha_\mathrm{P} + \alpha_\mathrm{F})
		\left( \gamma + \frac{4E}{3e} \right)
		- \frac{4E}{3e} (1 - 3 \alpha_\mathrm{P}) (1 + \alpha_\mathrm{F})}
		{1 + \alpha_\mathrm{P} + \alpha_\mathrm{F}
		+ 4 \alpha_\mathrm{P} \frac Ee (1 + \alpha_\mathrm{F})}
		\frac p\rho,
		\\
    	\frac{\delta E}{\delta \rho}
		=
		\frac{1 + 4 \alpha_\mathrm{P} \left(\gamma - 1 + \frac{4E}{3e}\right)}
		{1 + \alpha_\mathrm{P} + \alpha_\mathrm{F}
		+ 4 \alpha_\mathrm{P} \frac Ee (1 + \alpha_\mathrm{F})}
		\frac E\rho.
	\end{gather}
\end{subequations}
In conclusion, let us look at the limits of gas-pressure dominated and
radiation-pressure dominated discs, respectively. In the limit 
$\alpha_\mathrm{P} \rightarrow \infty$, $E/e \rightarrow 0$, 
suitable for a gas-pressure
dominated disc, we get
\begin{equation}
	\frac{\delta p}{\delta \rho}
	=
	\frac{\gamma + 4 \frac Ee \alpha_\mathrm{F}}
	{1 + 4 \frac Ee \alpha_\mathrm{F}}
	\frac p\rho, \quad
	\frac{\delta E}{\delta \rho}
	=
	0;
\end{equation}
and in this way recover our Eqs.~\eqref{soundspeed}.  For the case of a
radiation dominated disc where $\alpha_\mathrm{P} \rightarrow 0$,
\begin{equation}
  \ceff^2
  =
  \frac{\gamma p}{\rho}
  +
  \frac{
    4 \lambda E / 3 \rho 
  }{
  1 + \displaystyle \alpha_\mathrm{F}
  }
  .
\end{equation}
When using this value for $\ceff$, and setting $\lambda = 1/3$,
Eq.~\eqref{dispBS} becomes identical to the 
\cite{BlaesSocrates2001ApJ}
dispersion relation.  The general dispersion relation provided in this
appendix, thus, encompasses both the case of a gas-pressure dominated disc,
such as a protoplanetary disc, and the opposite extreme of a radiation-pressure
dominated system like an accretion disc around a black hole.
\bibliography{FlaigEtAl}

\begin{thebibliography}{}

\bibitem[\protect\citeauthoryear{{Armitage}}{{Armitage}}{1998}]{Armitage1998Ap%
J}
{Armitage} P.~J.,  1998, \apj, 501, L189+

\bibitem[\protect\citeauthoryear{Balbus \& Hawley}{Balbus \&
  Hawley}{1991}]{BalbusHawley1991ApJ}
Balbus S.,  Hawley J.,  1991, \apj, 376, 214

\bibitem[\protect\citeauthoryear{{Balbus}}{{Balbus}}{2003}]{Balbus2003ARAnA}
{Balbus} S.~A.,  2003, \araa, 41, 555

\bibitem[\protect\citeauthoryear{Balbus \& Hawley}{Balbus \&
  Hawley}{2006}]{BalbusHawley2006ApJ}
Balbus S.~A.,  Hawley J.~F.,  2006, \apj, 652, 1020

\bibitem[\protect\citeauthoryear{{Balsara} \& {Spicer}}{{Balsara} \&
  {Spicer}}{1999}]{BalsaraSpicer1999JcP}
{Balsara} D.~S.,  {Spicer} D.~S.,  1999, Journal of Computational Physics, 149,
  270

\bibitem[\protect\citeauthoryear{{Blaes}, {Hirose} \& {Krolik}}{{Blaes}
  et~al.}{2007}]{BlaesEtAl07}
{Blaes} O.,  {Hirose} S.,    {Krolik} J.~H.,  2007, \apj, 664, 1057

\bibitem[\protect\citeauthoryear{{Blaes} \& {Socrates}}{{Blaes} \&
  {Socrates}}{2001}]{BlaesSocrates2001ApJ}
{Blaes} O.,  {Socrates} A.,  2001, \apj, 553, 987

\bibitem[\protect\citeauthoryear{{Blaes} \& {Balbus}}{{Blaes} \&
  {Balbus}}{1994}]{BlaesBalbus1994ApJ}
{Blaes} O.~M.,  {Balbus} S.~A.,  1994, \apj, 421, 163

\bibitem[\protect\citeauthoryear{Brandenburg}{Brandenburg}{2008}]{Brandenburg2%
008PS}
Brandenburg A.,  2008, Physica Scripta, p. 014016

\bibitem[\protect\citeauthoryear{{Brandenburg}, {Nordlund}, {Stein} \&
  {Torkelsson}}{{Brandenburg} et~al.}{1995}]{BrandenburgEtAl1995ApJ}
{Brandenburg} A.,  {Nordlund} A.,  {Stein} R.~F.,    {Torkelsson} U.,  1995,
  \apj, 446, 741

\bibitem[\protect\citeauthoryear{Courant, Friedrichs \& Lewy}{Courant
  et~al.}{1928}]{CourantFriedrichLewy1928}
Courant R.,  Friedrichs K.,    Lewy H.,  1928, Math. Ann., 100, 32

\bibitem[\protect\citeauthoryear{{Fromang} \& {Nelson}}{{Fromang} \&
  {Nelson}}{2006}]{FromangNelson06}
{Fromang} S.,  {Nelson} R.~P.,  2006, \aap, 457, 343

\bibitem[\protect\citeauthoryear{{Fromang} \& {Papaloizou}}{{Fromang} \&
  {Papaloizou}}{2007}]{FromangPapaloizou07}
{Fromang} S.,  {Papaloizou} J.,  2007, \aap, 476, 1113

\bibitem[\protect\citeauthoryear{{Fromang}, {Papaloizou}, {Lesur} \&
  {Heinemann}}{{Fromang} et~al.}{2007}]{FromangEtAl2007}
{Fromang} S.,  {Papaloizou} J.,  {Lesur} G.,    {Heinemann} T.,  2007, \aap,
  476, 1123

\bibitem[\protect\citeauthoryear{{Hawley}}{{Hawley}}{2000}]{Hawley2000ApJ}
{Hawley} J.~F.,  2000, \apj, 528, 462

\bibitem[\protect\citeauthoryear{Hawley, Balbus \& Winters}{Hawley
  et~al.}{1999}]{HawleyEtAl1999ApJ}
Hawley J.~F.,  Balbus S.~A.,    Winters W.~F.,  1999, \apj, 518, 394

\bibitem[\protect\citeauthoryear{{Hawley}, {Gammie} \& {Balbus}}{{Hawley}
  et~al.}{1995}]{HawleyEtAl1995ApJ}
{Hawley} J.~F.,  {Gammie} C.~F.,    {Balbus} S.~A.,  1995, \apj, 440, 742

\bibitem[\protect\citeauthoryear{{Hirose}, {Krolik} \& {Stone}}{{Hirose}
  et~al.}{2006}]{HiroseEtAl06}
{Hirose} S.,  {Krolik} J.~H.,    {Stone} J.~M.,  2006, \apj, 640, 901

\bibitem[\protect\citeauthoryear{{King}, {Pringle} \& {Livio}}{{King}
  et~al.}{2007}]{KingEtAl07}
{King} A.~R.,  {Pringle} J.~E.,    {Livio} M.,  2007, \mnras, 376, 1740

\bibitem[\protect\citeauthoryear{{Kissmann}}{{Kissmann}}{2006}]{Kissmann2006}
{Kissmann} R.,  2006, PhD thesis, Ruhr-Universit\"at Bochum

\bibitem[\protect\citeauthoryear{{Kley} \& {Crida}}{{Kley} \&
  {Crida}}{2008}]{KleyCrida2008AnA}
{Kley} W.,  {Crida} A.,  2008, \aap, 487, L9

\bibitem[\protect\citeauthoryear{{Krolik}, {Hirose} \& {Blaes}}{{Krolik}
  et~al.}{2007}]{KrolikEtAl07}
{Krolik} J.~H.,  {Hirose} S.,    {Blaes} O.,  2007, \apj, 664, 1045

\bibitem[\protect\citeauthoryear{Kurganov, Noelle \& Petrova}{Kurganov
  et~al.}{2001}]{KurganovNoellePetrova2001}
Kurganov A.,  Noelle S.,    Petrova G.,  2001, SIAM J. Sci. Comput., 23, 707

\bibitem[\protect\citeauthoryear{{Levermore} \& {Pomraning}}{{Levermore} \&
  {Pomraning}}{1981}]{LevermorePomraning1981ApJ}
{Levermore} C.~D.,  {Pomraning} G.~C.,  1981, \apj, 248, 321

\bibitem[\protect\citeauthoryear{{Londrillo} \& {Del Zanna}}{{Londrillo} \&
  {Del Zanna}}{2000}]{LondrilloDelZanna2000}
{Londrillo} P.,  {Del Zanna} L.,  2000, \apj, 530, 508

\bibitem[\protect\citeauthoryear{{Miller} \& {Stone}}{{Miller} \&
  {Stone}}{2000}]{MillerStone00}
{Miller} K.~A.,  {Stone} J.~M.,  2000, \apj, 534, 398

\bibitem[\protect\citeauthoryear{{Papaloizou} \& {Nelson}}{{Papaloizou} \&
  {Nelson}}{2003}]{PapaloizouNelson03}
{Papaloizou} J.~C.~B.,  {Nelson} R.~P.,  2003, \mnras, 339, 983

\bibitem[\protect\citeauthoryear{{Pessah}, {Chan} \& {Psaltis}}{{Pessah}
  et~al.}{2007}]{PessahEtAl07}
{Pessah} M.~E.,  {Chan} C.-k.,    {Psaltis} D.,  2007, \apjl, 668, L51

\bibitem[\protect\citeauthoryear{{Pringle}}{{Pringle}}{1981}]{Pringle1981ARAnA}
{Pringle} J.~E.,  1981, \araa, 19, 137

\bibitem[\protect\citeauthoryear{Sano \& Inutsuka}{Sano \&
  Inutsuka}{2001}]{SanoInutsaka2001}
Sano T.,  Inutsuka S.-i.,  2001, The Astrophysical Journal Letters, 561, L179

\bibitem[\protect\citeauthoryear{Sano \& Miyama}{Sano \&
  Miyama}{1999}]{SanoMiyma99}
Sano T.,  Miyama S.~M.,  1999, The Astrophysical Journal, 515, 776

\bibitem[\protect\citeauthoryear{{Shakura} \& {Syunyaev}}{{Shakura} \&
  {Syunyaev}}{1973}]{ShakuraSyunyaev1973AnA}
{Shakura} N.~I.,  {Syunyaev} R.~A.,  1973, \aap, 24, 337

\bibitem[\protect\citeauthoryear{Turner}{Turner}{2004}]{Turner04}
Turner N.~J.,  2004, The Astrophysical Journal Letters, 605, L45

\bibitem[\protect\citeauthoryear{{Turner}, {Stone}, {Krolik} \&
  {Sano}}{{Turner} et~al.}{2003}]{TurnerEtAl03}
{Turner} N.~J.,  {Stone} J.~M.,  {Krolik} J.~H.,    {Sano} T.,  2003, \apj,
  593, 992

\bibitem[\protect\citeauthoryear{{Turner}, {Stone} \& {Sano}}{{Turner}
  et~al.}{2002}]{TurnerEtAl2002ApJ}
{Turner} N.~J.,  {Stone} J.~M.,    {Sano} T.,  2002, \apj, 566, 148

\end{thebibliography}
\label{lastpage}
\end{document}